\documentclass[aps,pre,a4paper,twocolumn,english,amsmath,showpacs,floatfix]{revtex4}
\usepackage[T1]{fontenc}
\usepackage[latin1]{inputenc}
\usepackage{graphicx}
\usepackage{epsfig}

\makeatletter

\usepackage{graphicx}

\usepackage{babel}
\makeatother
\begin{document}

\title{
Capillary ordering and layering transitions in two-dimensional hard-rod  
fluids}
\author{Yuri Mart\'{i}nez-Rat\'{o}n}
\email{yuri@math.uc3m.es}
\affiliation{Grupo Interdisciplinar de Sistemas Complejos, 
Departamento de Matem\'aticas,
Escuela Polit\'ecnica Superior, Universidad Carlos III de Madrid, 
Avenida de la Universidad 30, E--28911, Legan\'es, Madrid, SPAIN}
\date{\today}

\begin{abstract}
In this article we calculate the surface phase diagram of 
a two-dimensional hard-rod fluid confined 
between two hard lines. In a first stage we study the 
semi-infinite system consisting of an isotropic fluid in contact with 
a single hard line. We have found complete wetting by the columnar phase at the 
wall-isotropic fluid interface. When the fluid is confined between 
two hard walls, capillary columnar ordering occurs via a first-order phase 
transition. For higher chemical potentials 
the system exhibits layering transitions even for very narrow slits (near the 
one-dimensional limit). The theoretical model used was a density-functional 
theory based on the Fundamental-Measure Functional applied to a fluid of 
hard rectangles in the restricted-orientation approximation (Zwanzig model). 
The results presented here 
can be checked experimentally in two-dimensional granular media made of rods, 
where vertical motions induced by an external source and excluded 
volume interactions between the grains 
allow the system to explore those stationary states which
entropically maximize packing configurations.  
We claim that some of the surface phenomena found 
here can be present in two-dimensional granular-media fluids.
\end{abstract}


\maketitle

\section{Introduction}
\label{intro}

The effect of fluid confinement on phase transitions   
is nowadays an active line of scientific research due to the direct 
application of the theoretically predicted surface phase diagrams 
in the nanotechnology industry. Confining simple fluids, such as 
hard \cite{Schmidt,Dijkstra} or Lennard-Jones \cite{Binder} spheres, 
in a narrow slit geometry, results in a rich phase behavior, which has 
recently been studied in detail. Liquid crystals confined in 
nanopores is another typical example of systems with important applications in 
the industry of electronic devices. For this reason 
they have been extensively studied using theoretical models based 
on density-functional theory. In particular, capillary phase transitions 
exhibited by a nematic fluid confined between hard walls\cite{Roij,Harnau}, 
or walls favoring a particular anchoring \cite{Kike}, have been predicted. 
When non-uniform liquid-crystal phases, such as the smectic phase, are
included in the study of confined systems, the resulting surface phase 
diagrams display a rich phenomenology, which includes wetting transitions, 
the appearance of smectic defects \cite{Samo},  and    
layering transitions \cite{Dani}. 

The effect of confinement on two-dimensional fluids 
is also an interesting topic of research. Langmuir monolayers 
of lipids on the surface of water have been extensively studied in the
last hundred years \cite{Kaganer}, and the discovery of structures and phase 
transitions in these systems has experienced a
dramatic evolution driven by the new experimental techniques. 
Now it is possible to confine these two-dimensional systems 
by external potentials and study the 
influence of the confinement on the molecular packing of surface 
monolayers.   
  
Another paradigm of two-dimensional systems where the confinement plays 
an important role is the packing structures formed by particles in
granular media \cite{Aranson}. The crystallization of a quasi-two-dimensional 
one-component granular-disk fluid has recently been studied experimentally 
\cite{Reis}. It was found that the properties of the
crystal structure obtained (such as packing fraction, lattice structure, 
and Lindenman parameter) coincide with their   
counterparts obtained from MC simulations of a hard disk fluid. 
Recent experiments have found non-equilibrium steady states 
in a vibrated granular rod monolayer with tetratic, nematic and 
smectic correlations \cite{Narayan}. Some of these textures are 
similar to the   
equilibrium thermodynamic states of two-dimensional anisotropic 
fluids resulting from density-functional calculations \cite{Yuri1} 
and MC simulations \cite{Donev}. It was shown by several authors 
that the inherent states of some frozen granular systems can be described by 
equilibrium statistical mechanics \cite{Coniglio}. Also, an 
experimental test of the thermodynamic approach to granular media
has recently been carried out \cite{Dean}. 
Confining two-dimensional granular rods in different 
geometries (circular, rectangular, etc.) results in spontaneous formation
of patterns, with different orientationally ordered 
textures and defects next to the container \cite{Galanis}. The 
authors of Ref. \cite{Chaudhuri} have carried out MC 
simulations of a confined hard disk fluid. They have found that  
the crystal phase fails to
nucleate due to formation of smectic bands when the system is confined 
\cite{Chaudhuri}. It would be interesting to device an experiment with 
confined granular disks with the aim of comparing the 
properties of the non-uniform stationary 
states with those obtained from the statistical mechanics applied to the 
hard disk fluid. 

The main purpose of this article 
is the study of a confined two-dimensional hard-rod fluid. We are 
interested in the calculation of the surface phase diagram of a 
hard rectangle (HR) fluid confined by a single or two hard lines. We 
can think on a HR fluid as an experimental realization of a 
system of hard cylinders confined between two plates at a distance 
less than twice the cylinder diameter.   
We suggest that some of the surface phase transitions 
obtained here by applying the density 
functional formalism to a confined two-dimensional HR fluid should be similar  
to the steady states of confined granular rods. Some experiments are 
required to verify this hypothesis. 

The paper is organized as follows. In Sec. II we present the 
theoretical model: the fundamental-measure density functional 
applied to a HR fluid in the restricted-orientation approximation. 
This section is divided into two subsections. In the 
first the model is particularized to the study of the bulk phases, 
while in the second part the  
theoretical expressions used in the calculations of the 
thermodynamic and structural properties of the 
interfaces are presented. 
The results are presented in Sec. III. First we study the bulk phase 
diagram of a HR fluid with aspect ratio equal to 3, and then the 
resulting surface phase diagrams of a single wall-HR fluid interface 
and of the fluid confined between two hard lines are presented. Some 
conclusions are drawn in Sec. IV.

\section{Theoretical model}
In this section we introduce the theoretical model used in the calculations 
of the bulk and interface equilibrium phases. To study highly inhomogeneous 
phases such as those resulting from the confinement of a fluid 
in a narrow slit geometry or the solid phase with a high packing fraction, 
we have used the Fundamental-Measure Theory (FMT) applied to an anisotropic 
fluid of hard rectangles. It is well known that this formalism presents
a great advantage over other techniques when dealing with highly 
inhomogeneous phases, and that this is mainly due to the fact that a basic
requirement to construct the FMT density functional is that it conform
with the dimensional cross-over criterium \cite{Yasha,Yuri}. 
To implement the calculations we have used the restricted-orientation
approximation, where the axes of the rectangles are restricted 
to align only along the coordinate axes $x$ or $y$.  
Thus, the whole system is described in terms of density profiles 
$\rho_{\nu}({\bf r})$ ($\nu=x,y$). 

While the ideal part of the free energy density in reduced 
thermal units has the exact form 
\begin{eqnarray}
\Phi_{\rm{id}}({\bf r})=\sum_{\nu}\rho_{\nu}({\bf r})\left[
\ln \rho_{\nu}({\bf r})-1\right],
\end{eqnarray} 
the FMT interaction part of the 2D HR fluid is approximated \cite{Yuri} by 
\begin{eqnarray}
\Phi_{\rm{exc}}({\bf r})=-n_0({\bf r})\ln\left[1-n_2({\bf r})\right]+
\frac{n_{1x}({\bf r})n_{1y}({\bf r})}{1-n_2({\bf r})},
\end{eqnarray}
where the weighted densities $n_{\alpha}$'s are calculated as 
\begin{eqnarray}
n_{\alpha}({\bf r})=\sum_{\nu=x,y}\left[\rho_{\nu}\ast \omega^{(\alpha)}
_{\nu}\right]({\bf r}),
\end{eqnarray}
and where the symbol $\ast$ stands for convolution, i.e., 
$\rho_{\nu}\ast \omega^{(\alpha)}_{\nu}=
\int_V d{\bf r}' \rho_{\nu}({\bf r}')\omega^{(\alpha)}_{\nu}
({\bf r}-{\bf r}')$. The weights $\omega^{(\alpha)}_{\nu}$ are 
the characteristic functions whose volume integrals constitute
the fundamental measures of a single particle (the 
edge lengths and surface area). They are defined as 
\begin{eqnarray}
\omega^{(0)}_{\nu}({\bf r})&=&\frac{1}{4}\delta(\frac{\sigma_{\nu}^x}{2}
-|x|)\delta(\frac{\sigma_{\nu}^y}{2}-|y|),\\
\omega^{(1x)}_{\nu}({\bf r})&=&\frac{1}{2}
\Theta(\frac{\sigma_{\nu}^x}{2}-|x|)\delta(\frac{\sigma_{\nu}^y}{2}-|y|),\\
\omega^{(1y)}_{\nu}({\bf r})&=&\frac{1}{2}
\delta(\frac{\sigma_{\nu}^x}{2}-|x|)\Theta
(\frac{\sigma_{\nu}^y}{2}-|y|),\\
\omega^{(2)}_{\nu}({\bf r})&=&\Theta(\frac{\sigma_{\nu}^x}{2}-|x|)
\Theta(\frac{\sigma_{\nu}^y}{2}-|y|),
\end{eqnarray}
where $\sigma_{\mu}^{\nu}=\sigma+(L-\sigma)\delta_{\mu\nu}$, with 
$L$ and $\sigma$ the length and width of the rectangle and $\delta_{\mu\nu}$ 
the Kronecker function, while 
$\delta(x)$ and $\Theta(x)$ are the Dirac delta and 
Heaviside functions, respectively.

\subsection{The bulk phases}
\label{Bulk}
To calculate the bulk phase diagram we need to minimize the 
Helmholtz free energy functional $\beta {\cal F}[\{\rho_{\nu}({\bf r})\}]=
\int d{\bf r} \left[\Phi_{\rm{id}}({\bf r})+\Phi_{\rm{exc}}({\bf r})
\right]$ with 
respect to the density profiles $\rho_{\nu}({\bf r})$. These density 
profiles have the symmetries corresponding to the equilibrium phases, which
can be the isotropic or nematic fluids, the smectic phase 
(with particles arranged in layers with their
long axes pointing perpendicular to the layers), 
the columnar phase (with long axes parallel to the layers), 
plastic solid (particles located at the nodes of the 
square grid with averaged orientational order parameter 
over the cell equal to zero), and oriented solid (with both translational 
and orientational order). To take proper account of all these 
possible symmetries, we have used a Fourier-series expansion of the density 
profiles:
\begin{eqnarray}
\rho_{\nu}({\bf r})=\rho_0 x_{\nu}\sum_{{\bf k}=(0,0)}^
{{\bf N}}\alpha_{k_1,k_2}^{(\nu)}\cos (q_1 x)\cos (q_2 y),
\end{eqnarray} 
where we defined ${\bf k}\equiv(k_1,k_2)$ [with ${\bf N}=N(1,1)$],  
$q_1=2\pi k_1/d_x$, and $q_2=2\pi k_2/d_y$ are the 
wave vector components parallel 
to $x$ and $y$ axes respectively, and $d_x$, $d_y$ are the periods of the 
rectangular grid along these directions. $\alpha_{k_1,k_2}^{(\nu)}$ are 
the Fourier amplitudes of the density profile of the species $\nu$ with 
the constraint $\alpha_{0,0}^{(\nu)}=1$. $\rho_0$ 
is the average of the local density over the cell, while $x_{\nu}$ is 
the cell-averaged occupancy probability of species $\nu$. The 
Fourier series is truncated at that value $N$ which guarantees that 
$\alpha_{N,N}^{(\nu)}<10^{-7}$. With this parametrization the 
weighted density can be calculated explicity as 
\begin{eqnarray}
n_{\alpha}({\bf r})=\rho_0\sum_{\nu,{\bf k}}x_{\nu}
\alpha_{k_1,k_2}^{(\nu)}\hat{\omega}^{(\alpha)}_{\nu}({\bf k})
\cos (q_1 x)\cos (q_2 y),
\end{eqnarray}
where $\hat{\omega}^{(\alpha)}_{\nu}({\bf k})$ are the Fourier transforms 
of the corresponding weights, which have the form
\begin{eqnarray}
\hat{\omega}_{\nu}^{(0)}({\bf k})&=&\chi_0(q_1\sigma_{\nu}^x/2)
\chi_0(q_2\sigma_{\nu}^y/2),\\
\hat{\omega}_{\nu}^{(1x)}({\bf k})&=&\sigma_{\nu}^x 
\chi_1(q_1\sigma_{\nu}^x/2)\chi_0(q_2\sigma_{\nu}^y/2),\\
\hat{\omega}_{\nu}^{(1y)}({\bf k})&=&\sigma_{\nu}^y
\chi_0(q_1\sigma_{\nu}^x/2)\chi_1(q_2\sigma_{\nu}^y/2),\\
\hat{\omega}_{\nu}^{(2)}({\bf k})&=&a 
\chi_1(q_1\sigma_{\nu}^x/2)\chi_1(q_2\sigma_{\nu}^y/2),
\end{eqnarray}
Here $a=L\sigma$ is the surface area of the particle, and $\chi_0(x)=\cos x$, 
$\chi_1(x)=\sin (x)/x$. We have selected the orientational director 
parallel to $y$. Thus, the equilibrium smectic (columnar) phase should be found 
by minimizing the free energy with respect to the Fourier 
amplitudes $\alpha_{0,k}^{(\nu)}$ ($\alpha_{k,0}^{(\nu)}$), 
the smectic (columnar) period $d_y$ ($d_x$) and the 
order parameter $Q_{\rm{N}}\in[-1,1]$ [related to the $x_{\nu}$'s 
through the relations $x_{\parallel,\perp}=(1\pm Q_{\rm{N}})/2$ where 
the symbols $\parallel,\perp$ stand 
for particle alignment along $y$ and $x$ respectively]. For uniform 
phases [$\alpha_{k_1,k_2}^{(\nu)}=0$ $\forall (k_1,k_2)\neq (0,0)$]
$Q_{\rm{N}}$ 
coincides with the nematic order parameter.  
The solid phase is to be found by minimizing the free energy with respect 
to all the Fourier amplitudes $\alpha_{k_1,k_2}^{(\nu)}$, the crystal periods 
$d_x$ and $d_y$, and the order parameter $Q_{\rm{N}}$ 
in the case of an orientationally ordered solid. To measure the packing 
structure and the orientational order of the bulk phases we use the 
local density and the order parameter profiles, 
$\rho({\bf r})=\sum_{\nu}\rho_{\nu}({\bf r})$, and 
$Q({\bf r})=\left[\rho_y({\bf r})-\rho_x({\bf r})\right]/\rho({\bf r})$ 
respectively.

\subsection{The interfacial phases}

As we want to study the hard wall-fluid interface or the HR fluid confined 
in a slit geometry, we have introduced the following external potential:
\begin{eqnarray}
V_{\nu}(x)=\left\{
\begin{array}{ll}
\infty, \quad  & x< \sigma_{\nu}^x/2  
\\\\ 0,  \quad & x \ge  \sigma_{\nu}^x/2,
\end{array}
\right.
\end{eqnarray}
for the semi-infinite system, and
\begin{eqnarray}
V_{\nu}(x)=\left\{
\begin{array}{ll}
\infty, \quad  & x< \sigma_{\nu}^x/2 \quad \text{and}\quad 
x> H-\sigma_{\nu}^x/2
\\\\ 0,  \quad & \sigma_{\nu}^x/2\le x \le  H-\sigma_{\nu}^x/2,
\end{array}
\right.
\end{eqnarray}
for the slit geometry, where $H$ is the slit width, and the normal 
to the wall was selected in the $x$ direction. Note that this 
external potential represents a hard line which excludes the  
center of mass of particles at distances less than their contact 
distances with the wall. 
In this sense we can say that the external potential favors 
parallel alignment at the wall. 
This is in contrast 
with the favored homeotropic 
alignment usually considered in several studies of three-dimensional 
liquid crystals confined by a single or two walls (in particular that of 
Ref. \cite{Dani}).

The one-dimensional equilibrium density profiles $\rho_{\nu}(x)$ were 
found by minimizing the excess surface free energy  per unit length
\begin{eqnarray}
\gamma\equiv \int dx\left\{\frac{\Phi(x)}{\beta}
+P-\sum_{\nu}\rho_{\nu}(x)
\left[\mu_{\nu}-V_{\nu}(x)\right]\right\},
\label{surface}
\end{eqnarray} 
where $\beta=(k_{B}T)^{-1}$, $\Phi(x)=\Phi_{\rm{id}}(x)+\Phi_{\rm{exc}}(x)$, 
and $\mu_{\nu}$ are the chemical potentials of species $\nu$ fixed 
at the bulk fluid-phase value at infinite distance 
from the wall, while $P$ is the fluid pressure. The chemical potential 
of the bulk fluid phase is calculated, as usual, as $\mu=\sum_{\nu} x_{\nu}
\mu_{\nu}$, with $x_{\nu}$ the molar fractions of species $\nu$. If the 
bulk phase is an isotropic fluid then $x_{\nu}=1/2$, and $\mu_{\nu}=\mu$, 
$\forall\nu$. 

To measure the degree of interfacial order, we will use the
adsorption of the density profile, defined as 
$\Gamma=\int dx \left[\rho(x)-\rho(\infty)\right]$, and the  
order parameter profile $Q(x)$. 

The expression (\ref{surface})   
coincides with the definition of the 
surface tension of the wall-fluid interface for the semi-infinite case, 
which is approximately equal to half the excess surface free-energy   
for the slit geometry when the wall distance $H$ is large enough to 
accommodate both interfaces. 

To minimize the functional given by (\ref{surface}), we have discretized
space in the $x$ direction and minimize $\gamma$ with respect to
$\rho_{\nu}(x_i)$ ($x_i\in [x_0,x_N]$) using the conjugate-gradient algorithm.

\section{Results}
In this section we present the main results obtained from the 
application of the theoretical model just described to the study of 
surface properties of a 2D HR fluid. Particles were chosen to have
aspect ratio $\kappa\equiv L/\sigma=3$. This aspect ratio is
chosen because one of the aims of the present work is the study of layered
phases confined by one or two walls. As we will show bellow for 
$\kappa=3$ the stable phase is the columnar layered phase. 

In the first subsection we will summarize the results obtained in the 
calculation of the bulk phase diagram of this system, while in the 
second subsection we will focus on the study of the surface phase 
diagram. 

\subsection{Bulk phase diagram}
\begin{figure}[h]
{\centering \resizebox*{8.5cm}{!}{\includegraphics{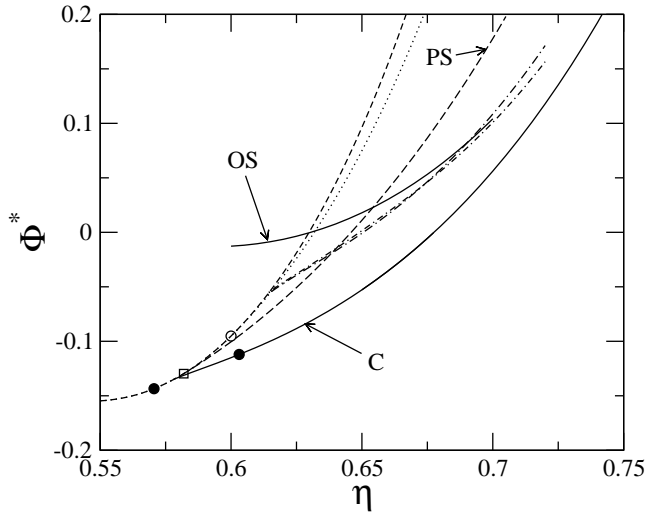}} \par}
\caption{\label{fig1}\small The rescaled 
free-energy density $\Phi^*=\Phi+2.9875-5.8501\eta$ is plotted against 
the mean packing fraction for all the stable and metastable phases found. These 
are: isotropic (dashed line), nematic (dotted line), smectic-1 and smectic-2 
(dotted and dashed lines), 
plastic solid (dashed line labelled as PS), while the perfectly oriented solid  
and the columnar phases (labelled in the figure as OS and C respectively) 
are plotted with solid lines. 
The open circle indicates the isotropic-nematic bifurcation point; 
the open square, 
the isotropic-plastic solid bifurcation point; 
and the solid circles represent the 
coexisting packing fractions at isotropic-columnar phase coexistence.}
\end{figure}

We have minimized the free energy density of the HR fluid,
defined as 
$\Phi\equiv V^{-1}\int_V d{\bf r}\left[\Phi_{\rm{id}}({\bf r})
+\Phi_{\rm{exc}}({\bf r})\right]$, 
with respect to the Fourier amplitudes, periods, and mean 
occupancy probability, as described in detail in Sec. \ref{Bulk}. The
results are plotted in Fig. \ref{fig1}, where the 
free-energy densities of all the stable and metastable phases found are 
plotted as a function of the packing fraction $\eta=\rho_0 a$. 
We have found, apart from the usual isotropic (I) and nematic (N) phases, 
two different smectic phases (Sm$_1$, and Sm$_2$), 
a plastic solid (PS), perfectly oriented solid (OS), and 
finally the columnar phase (C), which is the stable one in the whole range of 
packing fractions explored.

\begin{figure}[h]
{\centering \resizebox*{8.5cm}{!}{\includegraphics{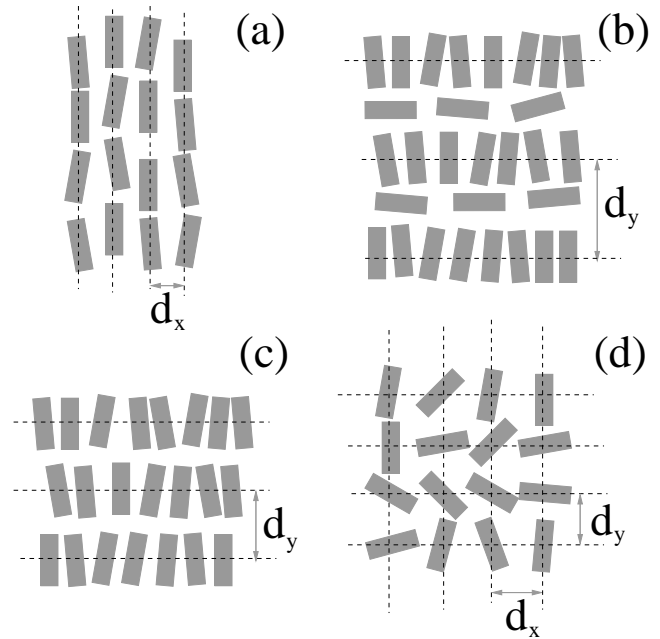}} \par}
\caption{\label{fig2}\small Sketch of particle configurations 
corresponding to different phases: columnar (a), smectic-1 
(b), smectic-2 (c), and plastic solid (d) phases. The direction of 
spatial periodicities are labeled in the figure.} 
\end{figure}

The coupling between the spatial and orientational
degrees of freedom of the particles results in the presence
of phases (stable or metastable) with different symmetries. In Fig.
(\ref{fig2}) we have sketched some of the particle configurations
corresponding to phases with columnar (a), smectic-1 (b), smectic-2 (c), and
plastic solid (d) symmetries found from the numerical minimization of
the density functional. The directions of spatial periodicities of
each phase have been depicted in the figure.

\begin{figure}[h]
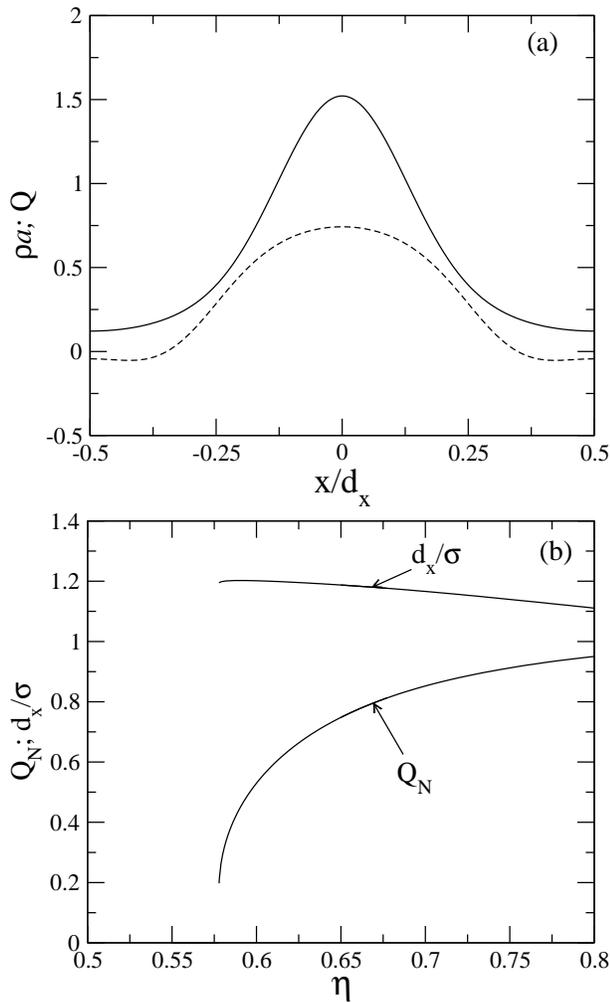

{\centering \resizebox*{8.cm}{!}{\includegraphics{fig3a.eps}} \par}
{\centering \resizebox*{8.cm}{!}{\includegraphics{fig3b.eps}} \par}
\caption{\label{fig3}\small (a): density $\rho(x)$ (solid line) 
and order parameter 
$Q(x)$ (dashed line) profiles of the columnar phase 
at coexistence with the isotropic phase. 
(b): order parameter $Q_{\rm{N}}$ and period of the columnar phase 
against the mean packing fraction.} 
\end{figure}

In Fig. \ref{fig3} (a) we have plotted the density and order-parameter 
profiles of the coexisting columnar phase. The columnar phase is 
orientationally ordered in the $y$ direction with the long rectangle 
axis pointing along the $y$ axis, while the periodicity of 
both density and order parameter profiles (which are in phase) 
is along the $x$ direction [see Fig. \ref{fig3} (a)].  
The mean coexistence
packing fractions of the I and C phases are $\eta_{\rm{I}}=0.57058$ and 
$\eta_{\rm{C}}=0.60310$, respectively while the period of the C phase,  
in units of the HR width, was found to be 
$d_x/\sigma=1.20102$. In Fig. \ref{fig3} (b)
we have plotted the order parameter $Q_{\rm{N}}$, 
and the period of 
the columnar phase as a function of the mean packing fraction. 

\begin{figure}[h]
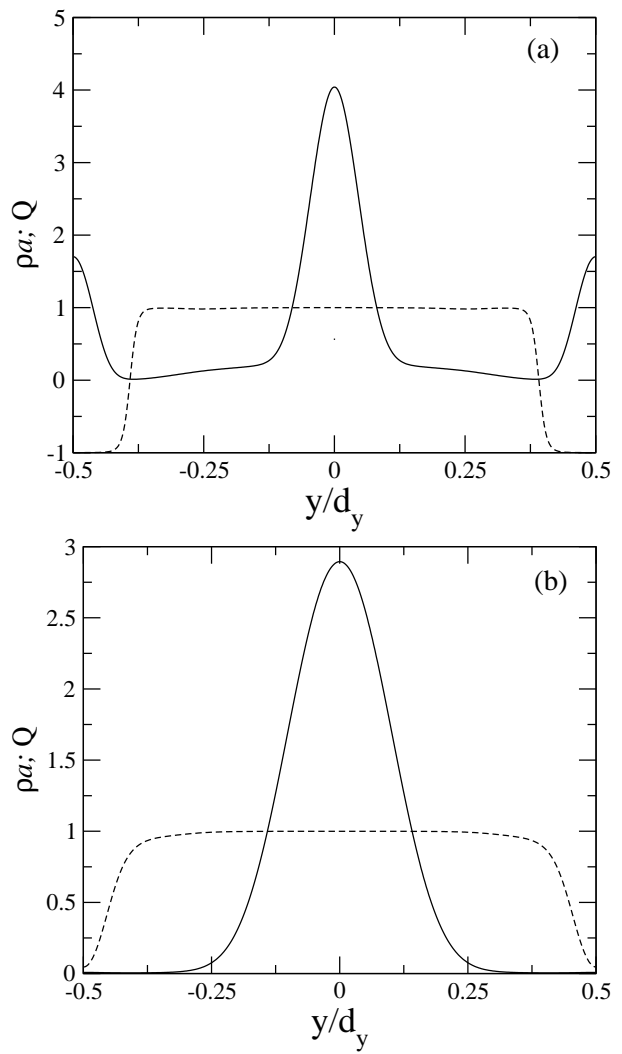

{\centering \resizebox*{8.cm}{!}{\includegraphics{fig4a.eps}} \par}
{\centering \resizebox*{8.cm}{!}{\includegraphics{fig4b.eps}} \par}
\caption{\label{fig4}\small 
Density (solid line) and order parameter (dashed line) 
profiles of the smectic-1 (a) 
and smectic-2 (b) metastable phases for a value of mean packing 
fraction fixed at 0.7.}
\end{figure}

\begin{figure}[h]
\epsfig{file=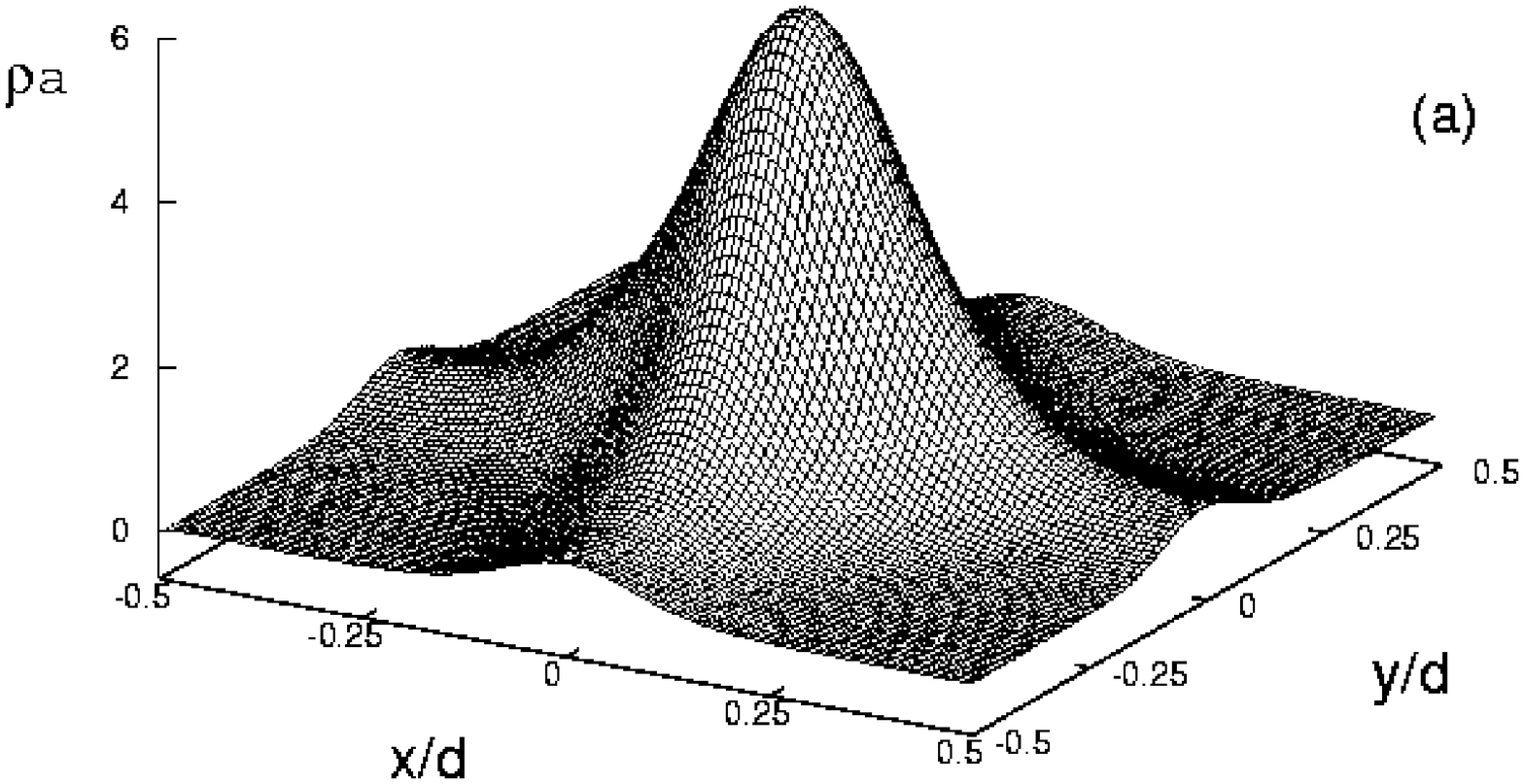,width=2.5in,angle=-90}
\epsfig{file=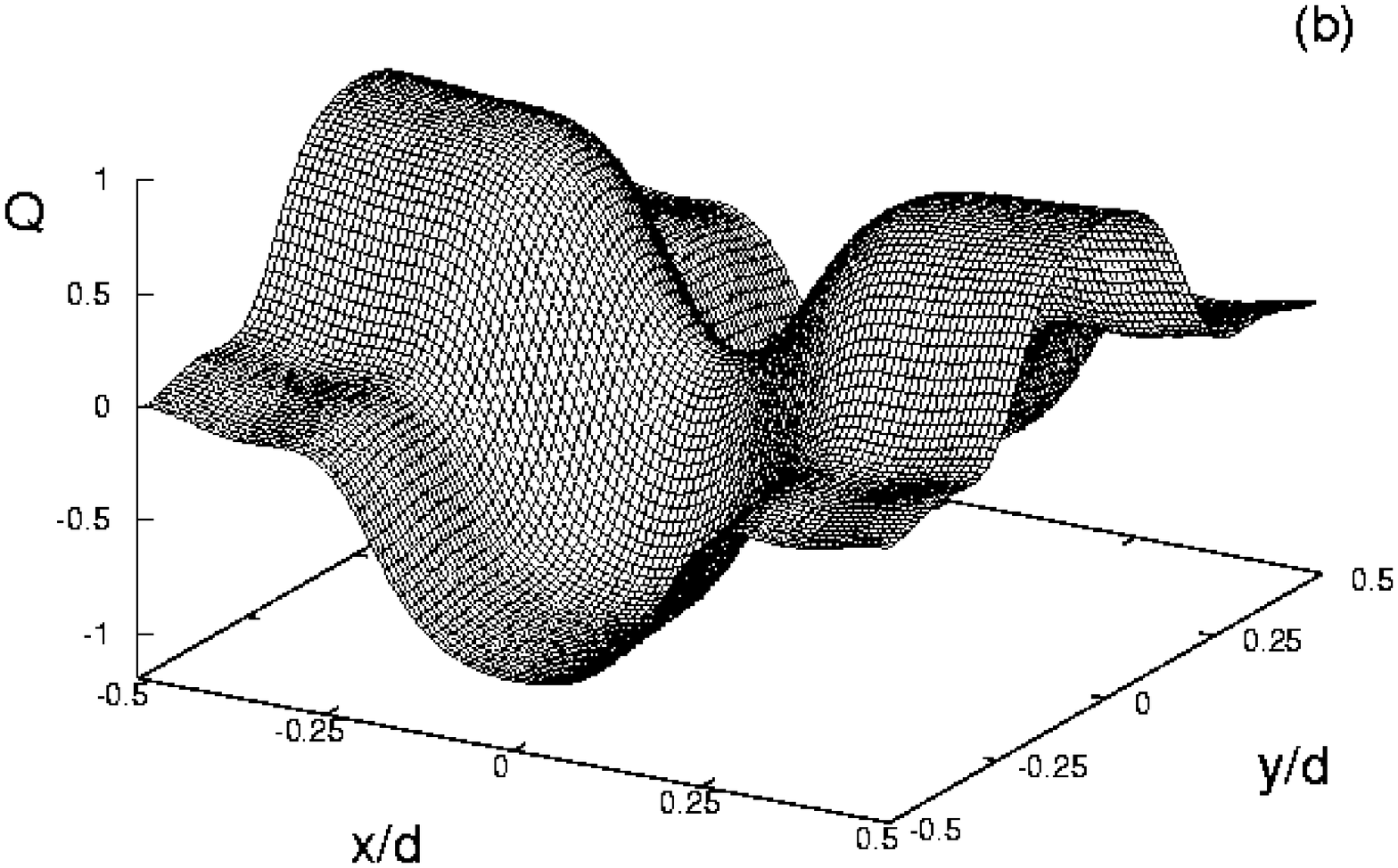,width=2.5in,angle=-90}
\epsfig{file=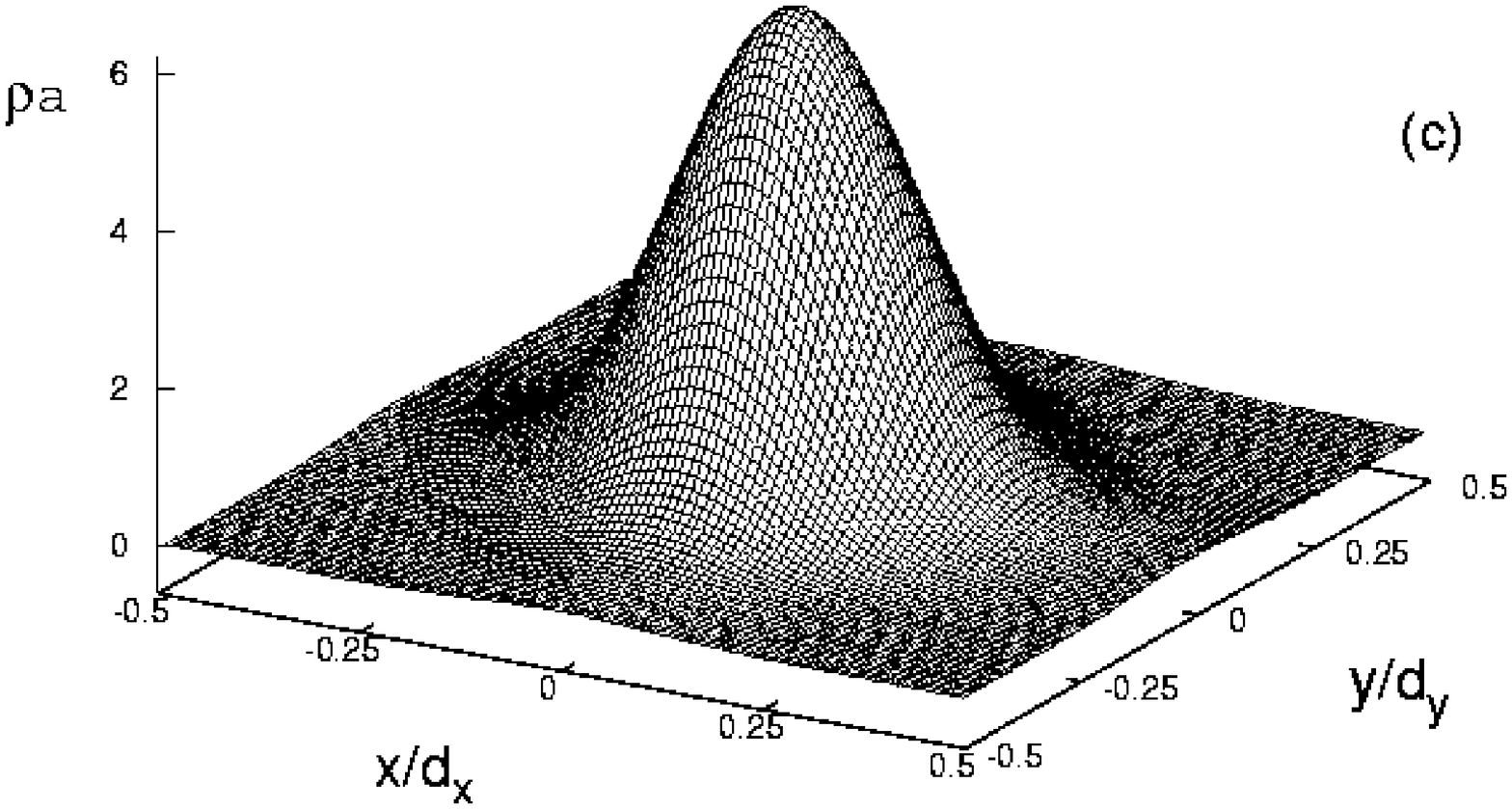,width=2.5in,angle=-90}
\caption{Density (a) and order-parameter (b) profiles of the plastic solid
 phase. (c):
Density profile of the perfectly oriented solid.} 
\label{fig5}
\end{figure}

To compare the different packings of HR particles in the
metastable phases (found as the local minima of the free energy density) for 
a fixed mean packing fraction $\eta=0.7$, we 
have plotted the density and order-parameter profiles 
of the Sm$_{1,2}$ [Fig. \ref{fig4} (a) 
and (b)], and PS and OS [Fig. \ref{fig5} (a)--(c)] phases. 
As can be seen from Fig. \ref{fig4} (a), the density profile of the 
Sm$_1$ phase has two maxima per period. The less pronounced 
maxima, located at the interstitials, reflect the high population of 
particles with 
long axes oriented parallel to the smectic layers [see the sketched 
particle configurations in Fig. 
\ref{fig2} (b)]. This alignment  
is also shown in the order-parameter profile, which reaches high negative 
values at the interstitial positions. This phase bears a strong 
resemblance to the findings of Refs. \cite{Roij2} and \cite{Allen} where 
the particle equilibrium configurations in the 3D smectic phases show 
the same pattern. As a consequence of this
(alternating population of particles aligned perpendicular  
--sharpest peak in the density profile-- and parallel to the layers), 
the smectic period in units of the particle length is $d_y/L=1.53025$,
higher than the smectic period of the Sm$_2$ phase 
($d_y/L=1.17935$). The density and 
order-parameter profiles of the Sm$_2$ are shown in Fig. \ref{fig4} (b). 
As can be seen from the figure, these profiles reflect the usual packing 
in smectics, characterized by a single density peak with vanishingly 
small population of particles in the interstitials, while the order 
parameter reaches its maximum value at the position of the 
smectic layers [see Fig. \ref{fig2} (c) for the sketched particle 
configurations].

The density and order parameter profiles of the PS phase with mean 
packing fraction equal to 0.7 are plotted in Fig. \ref{fig5} (a) and (b). 
The plastic solid has the same periodicity in the $x$ and $y$ direction, i.e. 
$d_x=d_y=d$, and the order parameter averaged over the unit cell is strictly 
equal to zero. As we can see from Fig. \ref{fig5} (b), while the order 
parameter at the nodes of the square lattice is equal to zero, it reaches 
positive (negative) values at the $(\pm 0.5,0)$ [$(0,\pm 0.5)$] positions 
along the sides of the cell (the same solution with the $x$ and $y$ directions 
interchanged was found in the minimization of the free energy). Finally, 
the density profile of the perfectly aligned two-dimensional solid is 
plotted in Fig. \ref{fig5} (c).  

Although the phases described above are metastable with respect to the 
columnar phase, they can be stabilized for different values of the 
particle aspect ratio. A detailed study of the complete phase diagram, 
necessary to elucidate this point, is a work in progress. 

We now proceed to make a comparison between the results
for the 2D Zwanzig model with $\kappa=3$ obtained above and
those for hard parallelepipeds with restricted orientations
and the same value of $\kappa$ \cite{PRE_Yuri}. This comparison will show
the differences in phase behaviour between three and two dimensions as
predicted by Fundamental-Measure Theory (which, as already pointed out,
conforms with the dimensional crossover criterion).
As shown in Ref. \cite{PRE_Yuri}, hard parallelepipeds
exhibit a second-order phase transition
between isotropic and plastic solid phases. As density increases
the system goes to a discotic smectic phase (confirmed by simulations)
via a first-order phase transition, which in turn discontinuously changes
to a columnar phase and then to an oriented solid. By contrast, the present
model shows that, in two dimensions, the isotropic phase exhibits a
first-order transition to a columnar phase that is stable until
very high packing fractions (more stable that plastic, oriented solid
and different smectic phases). As a consequence, one expects that
the corresponding surface phase diagrams be also different. 

\subsection{Surface phase diagram}
In this section we deal with surface phenomena. In 
the first part we will concentrate on the semi-infinite 
wall-isotropic interface of a HR fluid, while in the second part we will 
focus on the slit geometry. We will demostrate the presence of complete 
wetting, capillary ordering and layering transitions 
in the confined two dimensional
hard rod fluid. For a detailed discussion on general grounds of the  
phase behavior and critical phenomena 
of a confined by a single wall fluid see Ref. 
\cite{Lipowsky}. 

\subsubsection{The wall-fluid interface}
The interaction between the isotropic fluid phase and a hard wall was 
studied by calculating the one-dimensional 
equilibrium density $\rho(x)$ and order-parameter $Q(x)$ profiles through 
the excess surface free-energy minimization [see Eq. (\ref{surface})]. The 
chemical potential $\mu$ of the fluid phase at infinite distance 
from the wall was varied within the range of isotropic-phase stability, i.e. 
$\mu\in[-\infty,\mu_0]$ ($\mu_0$ being the value at which the 
I-C phase transition occurs). It is well known that the presence of a 
hard wall in a system of elongated particles induces parallel alignment 
of the particle axes with respect to the wall \cite{Holyst,Chrzanowska}.
This preferential alignment is a result of 
the entropic depletion effect. In the parallel configuration, the 
centers of mass of the particles are much closer to the wall, so the gain 
in volume per particle is larger and, as a consequence, the configurational 
entropy of the system is maximized. This effect is responsible for the 
occurrence of a biaxial nematic phase which breaks the orientational symmetry 
in a three-dimensional nematic fluid \cite{Roij}. The same depletion 
mechanism is at work in 2D, as we will show below.  

\begin{figure}[h]
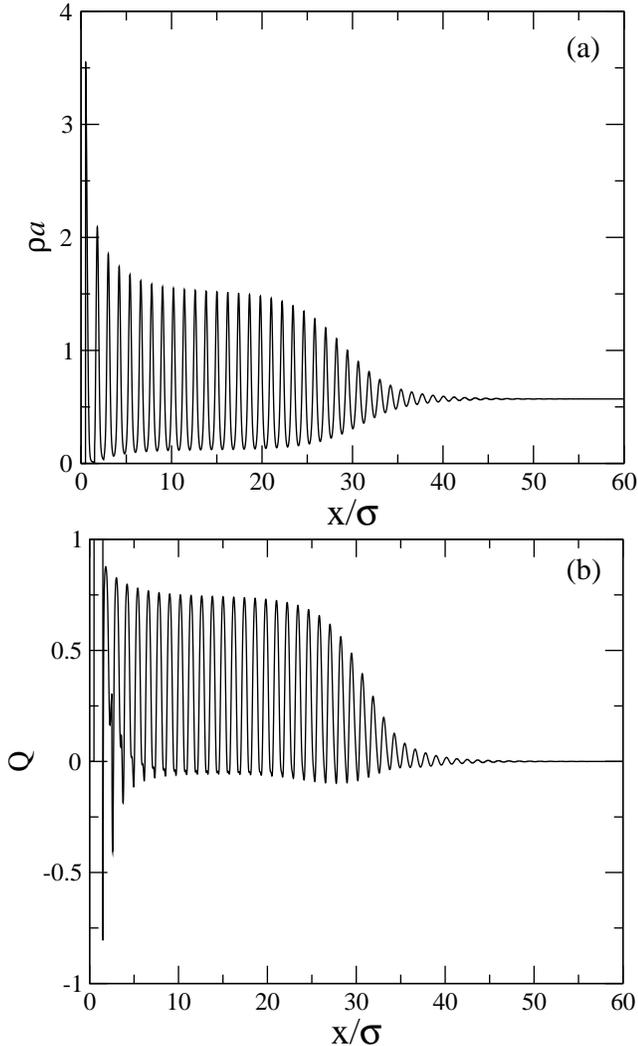

\epsfig{file=fig6a.eps,width=3.2in}
\hspace*{-0.4cm}
\epsfig{file=fig6b.eps,width=3.3in}
\caption{Density (a) and order parameter (b) profiles of the wall-isotropic 
fluid interface. The undersaturation is fixed to $\beta\Delta\mu=-1.1\times
10^{-4}$.}
\label{fig6}
\end{figure}

\begin{figure}[h]
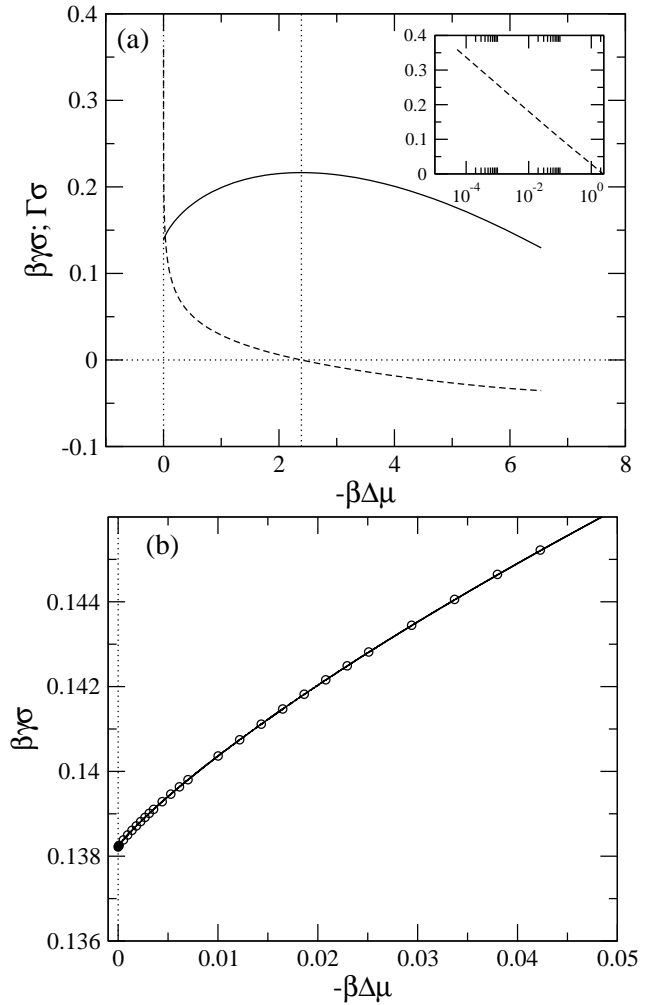

\epsfig{file=fig7a.eps,width=3.2in}
\hspace*{-0.2cm}
\epsfig{file=fig7b.eps,width=3.3in}
\caption{(a): excess surface free energy (solid line) and adsorption 
coefficient (dashed line), in reduced units, against $\beta\Delta\mu$. 
The inset shows $\Gamma \sigma$ vs. $\beta\Delta\mu$ in logarithmic scale. 
(b): excess surface free energy 
vs. $\beta\mu$ in the neighborhood of zero undersaturation. The open 
circles show the values obtained from the numerical 
minimization, while the solid line represents the analytic curve obtained 
by integrating the interfacial Gibbs-Duhem relation with the fitted 
logarithmic dependence of the adsorption coefficient (see text). 
The solid circle shows the value of the W-I surface tension 
$\beta\gamma_{\rm{WI}}\sigma=0.13822$}
\label{fig7}
\end{figure}

The results from the minimization are shown in Figs. \ref{fig6} (a) and (b) 
for an undersaturation of $\beta\Delta\mu=-1.1\times 10^{-4}$. 
As we can see from the figure, the density and order-parameter profiles 
indicate columnar order near the wall, which propagates several columnar periods
into the fluid phase. The wall-fluid interaction enhances the 
orientational order 
near the surface and the adsorption of particles, creating a structured layer 
with columnar-phase symmetry which grows in width 
with increasing chemical potential and diverges  
at $\mu=\mu_0$. Thus, complete wetting by a columnar 
phase occurs at the wall-isotropic interface. This result is 
shown in Fig. \ref{fig7} (a) where the excess surface free-energy $\gamma$ 
and the adsorption coefficient $\Gamma$ 
are plotted against $\Delta\mu=\mu-\mu_0$. As 
we can see, $\Gamma$ grows continuously, ultimately
diverging logarithmically with $\Delta\mu$ (see inset of 
figure). The excess surface free energy $\gamma$ has a maximum, and 
at this point the adsorption passes through zero. This result 
is directly related to the interfacial Gibbs-Duhem equation,
$\Gamma=-d\gamma/d\mu$, which relates the adsorption coefficient with 
the first derivative of the excess surface free-energy with respect to 
bulk chemical potential. At $\mu_0$ the excess surface energy is equal
to the wall-isotropic surface tension $\gamma_{\rm{WI}}$, which is in turn
equal to the sum of wall-columnar and columnar-isotropic surface 
tensions, $\gamma(\mu_0)=\gamma_{\rm{WI}}=\gamma_{\rm{WC}}+\gamma_{\rm{CI}}$ 
(the Young's equation for complete wetting). 
  
We have carried out a logarithmic fit of the adsorption coefficient 
with respect to undersaturation $\beta\Delta\mu=\beta\left(
\mu-\mu_0\right)$, and we find that 
$\Gamma\sigma=\tau_1+\tau_2\ln\left[\beta|\Delta\mu|\right]$, with 
$\tau_1=0.02387$ and $\tau_2=-0.03396$. Then, integrating the 
interfacial Gibbs-Duhem relation $\Gamma=-d\gamma/d\mu$, we  
find the expression 
\begin{eqnarray}
\beta\gamma\sigma\approx\beta\gamma_{\rm{WI}}\sigma-
\left[\tau_1+\tau_2\left(\ln\left(\beta|\Delta\mu|\right)-1\right)\right]
\beta\Delta\mu,
\end{eqnarray}
which approximates the excess surface free energy near complete 
wetting. The above expression is plotted against 
$\beta\Delta\mu$ in Fig. \ref{fig7} (b), where the results from direct 
calculation of $\beta\gamma\sigma$, using the equilibrium density profiles 
obtained, are also plotted. As we can see the agreement 
is excellent even for relatively high values of undersaturation. 

\begin{figure}[h]
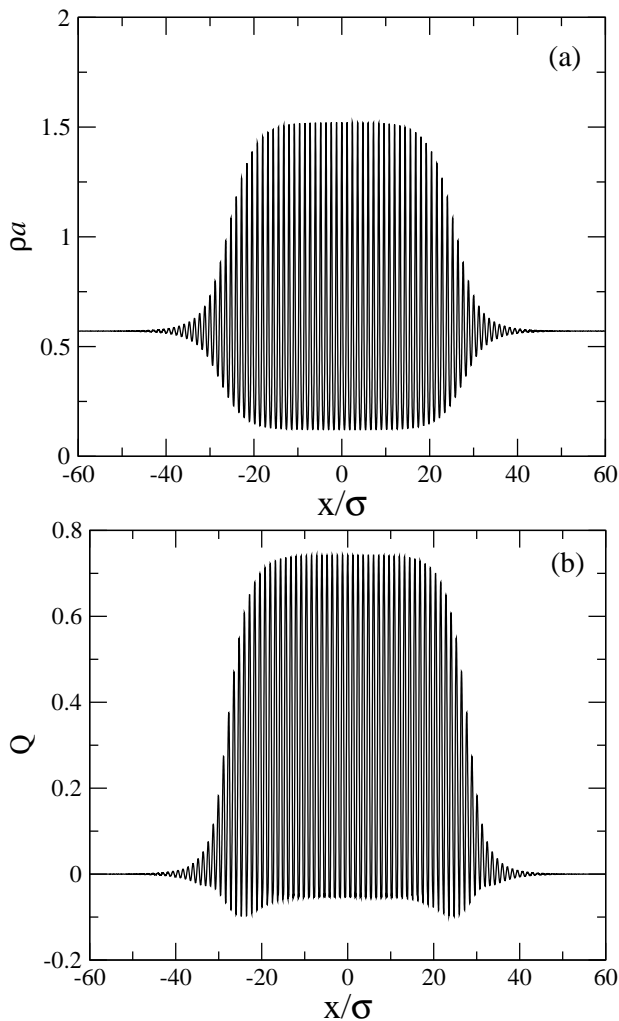

\epsfig{file=fig8a.eps,width=3.2in}
\epsfig{file=fig8b.eps,width=3.2in}
\caption{Density (a) and order-parameter (b) profiles of a numerical 
box containing two isotropic-columnar interfaces.}
\label{fig8}
\end{figure}

To calculate the structural and thermodynamic properties of the 
columnar-isotropic interface, we have implemented a numerical 
scheme already used in Ref. \cite{Yuri2}, consisting of
minimizing the surface excess free energy $\gamma$ 
in a box of width $h$ containing a stripe of a few columnar layers surrounded 
by isotropic material with periodic boundary conditions. $h$ is chosen 
such that the density profiles can easily accommodate the two interfaces 
and go to the coexistence fluid density at the periodic boundary.
A typical result from this calculation is plotted in Fig. \ref{fig8} (a) 
and (b) for the density and order-parameter profiles, respectively. Thus, 
the I-C interfacial tension can be calculated as half the excess 
surface free energy resulting from the minimization. We have found 
a value of $\beta\gamma_{\rm{IC}}\sigma=0.00672$.

\begin{figure}[h]
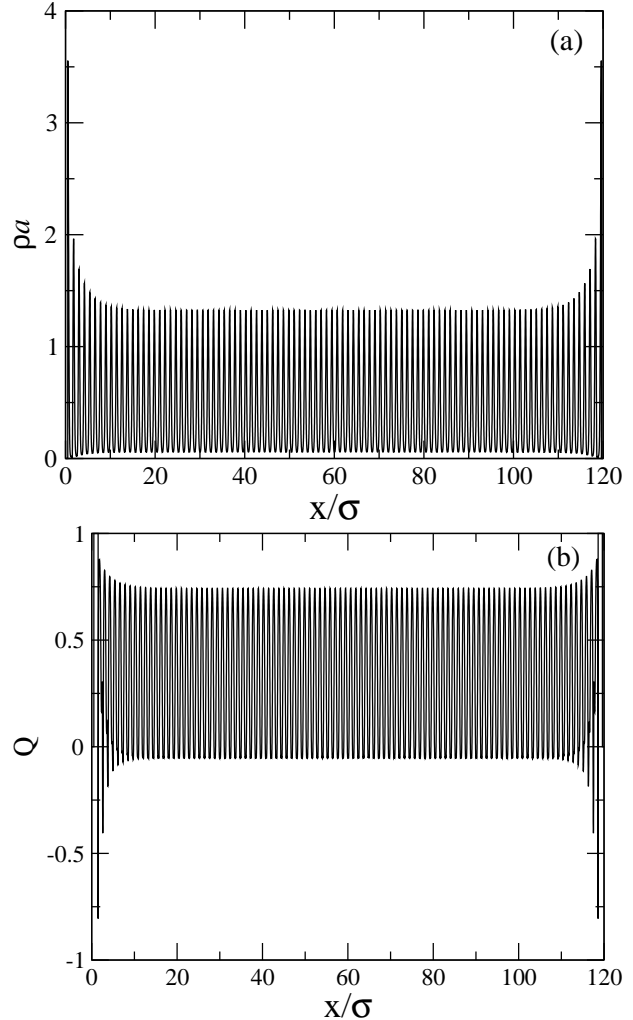

\epsfig{file=fig9a.eps,width=3.2in}
\epsfig{file=fig9b.eps,width=3.2in}
\caption{ The density (a) and order parameter (b) profiles of two 
wall-columnar interfaces}
\label{fig9}
\end{figure}

Finally, to verify that Young's law for complete wetting 
holds, we need to calculate the 
surface tension of the wall-columnar interface. To construct density 
profiles compatible with this semi-infinite interface, one has to
establish a boundary, at the side of the computational box opposite to the 
wall, and place, beyond the boundary and into the bulk, a
periodically structured profile, choosing the phase (i.e. the value
of the profile at the boundary) arbitrarily within the bulk period.
Although this recipe can in principle be implemented, we 
have chosen to fix bulk I-C coexistence conditions in a confined
columnar phase and calculate the density profile of the system
confined between two walls; the separation between the walls 
was chosen large enough so that the effects of having a finite 
interface penetration length caused by the presence of 
the confined external potential can be neglected. Also, in order
to ensure that commensurability effects can be ignored, the distance 
between the walls was set to a (large) integer number of equilibrium
periods of the columnar phase. The results from these calculations 
are plotted in Fig. \ref{fig9} (a) and (b). The W-C surface tension 
calculated as half the value of the excess surface free energy results 
in $\beta\gamma_{\rm{WC}}\sigma=0.13150$, compatible with Young's law 
in conditions of complete wetting of the W-I interface by the 
columnar phase. 

\subsubsection{Capillary ordering}

This section is devoted to a study of the effect of confinement 
of a 2D HR fluid on the thermodynamic and structural properties of the 
fluid. In particular, we are interested in the enhancement of the 
orientational and layering ordering due to confinement, 
and the commensurability effects exhibited by a layered phase 
sandwiched between two hard walls at a distance that may or may not
commensurate with the period of the bulk columnar phase. 
It is well known that, under certain circumstances (related to the nature
of the fluid-fluid and surface-fluid interactions), a fluid inside a pore 
can exhibit capillary first-order phase transitions between two different 
phases at a chemical potential below the bulk coexistence value. An 
example of this phenomenon is the recently studied capillary nematization 
\cite{Roij} and smectization \cite{Dani} of a liquid crystal fluid inside a 
pore. The bulk condensed phase may have uniform (nematic) or nonuniform 
density profiles. For the latter case, capillary layering transitions 
between interfacial phases with different number of smectic layers \cite{Dani}
can also be found. Here we will show that these capillary and layering 
phase transitions are not unique to 3D system. They are 
also present in 2D anisotropic fluids which can stabilize layered phases 
with different spatial symmetries, such as the columnar phase. 

\begin{figure}[h]
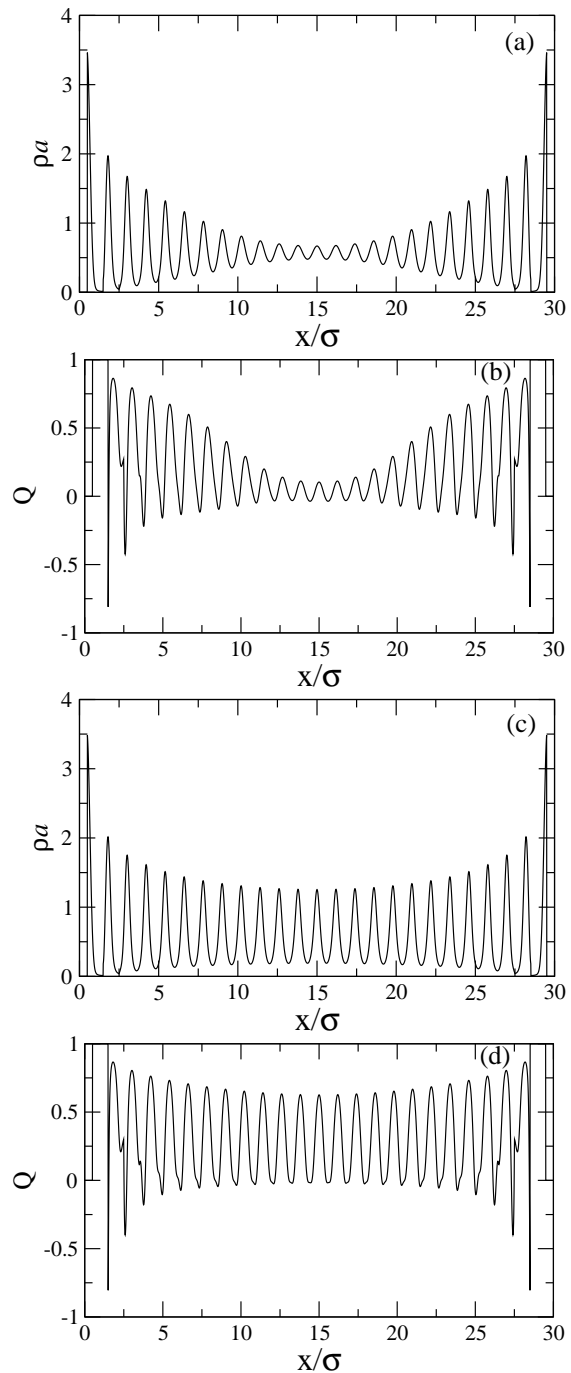

\epsfig{file=fig10a.eps,width=2.8in}
\hspace*{-0.4cm}
\epsfig{file=fig10b.eps,width=2.9in}
\epsfig{file=fig10c.eps,width=2.8in}
\hspace*{-0.4cm}
\epsfig{file=fig10d.eps,width=2.9in}
\caption{Isotropic (a)-(b) and columnar (c)-(d) phases that
coexist at the same chemical potential bellow $\mu_0$. 
(a), (c): Density profiles. 
(b), (d): Order-parameter profiles.}
\label{fig10}
\end{figure}

With a view to finding the effects of confinement on 
columnar ordering in a HR fluid, we have minimized the excess surface free 
energy with respect to the density profile for the particular case 
of HR's with $\kappa=3$. The fluid is confined by  
two hard walls at a distance $H/\sigma=30$ (in units of the particle width). 
As already pointed out, hard walls favor alignment parallel to the wall, 
as well as adsorption of particles at both surfaces (density and order 
parameters at contact are much higher than their bulk values). 
This coupled translational-orientational ordering near the surfaces 
propagates into the fluid, creating columnar ordering. We have found 
that for low values of the chemical potential of the bath the density
profile is structureless (except just at the wall contact), similar to the
bulk isotropic phase. Increasing the chemical potential several damped columnar
peaks appear near the wall in a continuous fashion, i.e. with their heights
increasing continuously.
At some value of 
the chemical potential, the system exhibits a first-order phase transition 
between a phase with highly damped columnar peaks to a new phase with much 
stronger columnar ordering even at the center of the pore. The typical 
density and order-parameter profiles of both interfacial phases 
are shown in Fig. \ref{fig10} (a)-(d). Although the less-ordered phase 
exhibits strong oscillations in both density and order-parameter profiles, 
the peak amplitudes are damped into the pore faster than those of 
the higher ordered phase. We will take the convention to 
call the first `isotropic', and the second `columnar' surface phases. This 
convention is justified by the fact that, just before the transition
described above, columnar ordering increases continuously, starting
from an isotropic-like density profile, as
the chemical potential is increased. Thus we cannot trace
out a definite boundary (a value for $\mu$ below that
corresponding to first order phase transition) below or above which
the profile inside the pore can be considered isotropic or
columnar. Only the first order
phase transition described above can really distinguish two
different surface phases, one of them less ordered (following our
convention, the isotropic phase) than the other (the columnar phase).
As we can see in the figure, the latter has 25 columnar peaks. 

\begin{figure}[h]
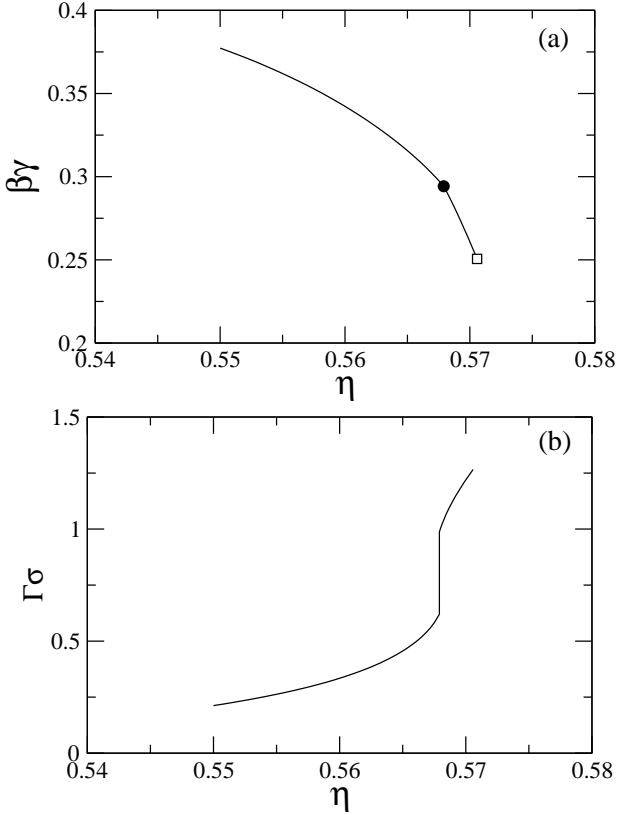

\hspace*{-0.3cm}
\epsfig{file=fig11a.eps,width=3.2in}
\epsfig{file=fig11b.eps,width=3.1in}
\caption{Excess surface free energy (a) and adsorption 
coefficient (b) against packing fraction of 
the bulk isotropic fluid. In the figure at top, the solid circle
represents the transition point between both interfacial phases, 
while open square indicates the point corresponding 
to the bulk coexistence value for isotropic and columnar phases.}
\label{fig11}
\end{figure}

The transition point is calculated from the discontinuity in 
the first derivative of the excess surface free energy with 
respect to the bulk packing fraction $\eta$. The corresponding plot is shown 
in Fig \ref{fig11} (a). At 
this point the adsorption coefficient jumps discontinuously from the less- 
(the damped columnar) to the higher-ordered 
phase [see Fig. \ref{fig11} (b)]. 

This surface transition point is located below the bulk isotropic-columnar 
phase transition [see Fig. \ref{fig11} (a)], showing the presence of  
columnar-order enhancement in the pore. On further increasing the 
chemical potential up to a sufficiently high value (above $\mu_0$), we find a 
first-order layering transition between two interfacial columnar phases 
which differ by just a single columnar layer. The behavior of the excess 
surface free energy and the adsorption coefficient 
is similar to that shown in Fig. \ref{fig11} (a) and (b). 
Alternatively we can find the transition from $n-1$ to $n$ columnar layers 
by fixing the chemical potential and increasing the pore width $H$. 

\begin{figure}[h]
\epsfig{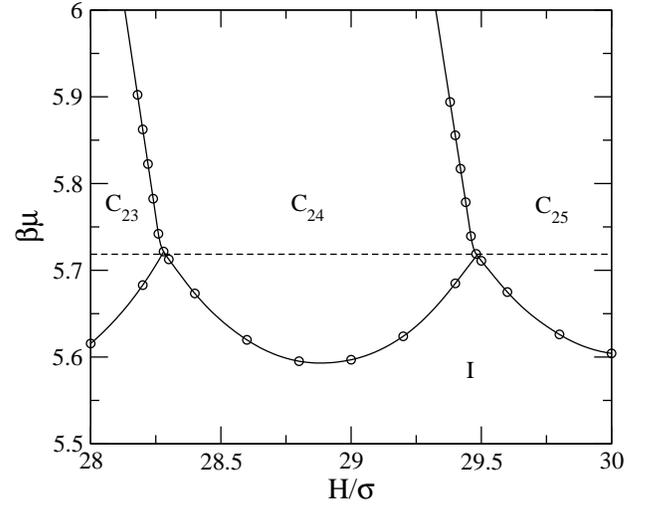}
\caption{$\mu-H$ surface phase diagram showing first-order 
capillary columnar ordering and layering transitions. The 
pore width covers a range which goes  
from 23 to 25 columnar layers, as labeled in the figure. The open circles 
indicate the cases chosen for calculations,
while the solid line is a cubic spline interpolation. 
The horizontal dashed line shows the value of the bulk 
chemical potential at the isotropic-columnar phase coexistence.} 
\label{fig12}
\end{figure}

The two surface phase transitions described above, namely
first-order capillary I-C ordering and $(n-1)$--$n$ layering transition,
are connected in the $\mu-H$ surface phase diagram through the peculiar 
structure shown in Fig. \ref{fig12}.

The parabola below the bulk transition line corresponds to first-order 
transition lines separating regions of stability of the isotropic and 
the columnar interfacial phases, while the 
straight lines indicate layering transitions. Increasing the chemical 
potential from low values to those corresponding to the parabola, the 
density profiles always change continously from a structureless to 
damped columnar density profile.   
Both types of  
transitions (the isotropic-columnar and $n-1$-$n$ layering transitions) 
coalesce in triple points, two of which are shown in Fig. 
\ref{fig12}. At the triple points an isotropic and two columnar 
interfacial phases with $n-1$ and $n$ layers coexist in equilibrium.
The set of connected  of Fig. \ref{fig12} are similar to those 
found in MC simulations of the confined hard-sphere fluid 
\cite{Dijkstra}. In this work the authors have shown the existence of capillary 
freezing of the HS fluid, confined in the slit geometry, for chemical 
potential values below the bulk freezing transition. The transitions lines 
in the $\mu-H$ surface phase diagram follow the same topology 
of connected set of parabolas as found in our system. 

Some of the topological features of this surface phase diagram 
can be elucidated from the 
Clausius-Clapeyron equation as applied to the interfacial coexistence lines. 
The excess surface free energy $\gamma(\mu,H)$ along coexistence 
is a function of two variables, the chemical potential $\mu$, and the pore 
width $H$. Thus, infinitesimal changes in these variables 
along the coexistence curve are related through the equation 
\begin{eqnarray}
d\gamma_{\alpha}-d\gamma_{\beta}=\Delta
\left(\frac{\partial \gamma}{\partial \mu}\right)_{H}d\mu
+\Delta\left(\frac{\partial \gamma}{\partial H}\right)_{\mu}dH=0,
\end{eqnarray}
where the coexisting condition $\gamma_{\alpha}=\gamma_{\beta}$ 
(for $\alpha,\beta=$I,C$_{n-1}$,C$_{n}$) was used, 
and $\Delta u=u_{\alpha}-u_{\beta}$ for any function $u$. Using 
the interfacial Gibbs-Duhem equation $\partial\gamma/\partial\mu=-\Gamma$ and 
the definition of the solvation force $f=-\partial\gamma/\partial H$, 
we arrive at
\begin{eqnarray}
\frac{d\mu}{dH}=-\frac{\Delta f}{\Delta \Gamma},
\label{res}
\end{eqnarray}
which relates the first derivative of the chemical potential 
with respect to the pore width with  
changes in the solvation force and in the adsorption coefficient  
at the transition point. The negative slope of the layering curves 
is a direct result of Eq. (\ref{res}), as the increment in the adsorption 
is always positive for the $(n-1)\to n$ layering transition, while
the change in the solvation force is also positive (the latter
can be interpreted as an increment with respect to the bulk of 
the excess surface pressure, which is obviously larger for the phase
with $n$ layers). For values of the pore width 
that commensurate with an integer number of columnar periods 
of the bulk columnar phase, the solvation force 
becomes zero and we get a minimum in 
the I-C capillary transition curve (see Fig. \ref{fig12}). At each 
side of the minimum the solvation force change the sign to positive 
(left side) or negative (right side) as we compress or expand the 
film, respectively, while the change in adsorption remains positive.  

The Kelvin equation for capillary condensation in a slit geometry  
relates the undersaturation in chemical potential with
pore width $H$ as 
\begin{eqnarray}
\Delta\mu=\mu(H)-\mu_0=-\frac{2\gamma_{\alpha\beta}}
{(\rho_{\alpha}-\rho_{\beta})H},  
\label{Kelvin}
\end{eqnarray}
where $\rho_{\alpha}$ and $\rho_{\beta}$ are the bulk 
coexisting densities of phases $\alpha$ and $\beta$ ($\alpha$ 
being the condensed phase), while 
$\gamma_{\alpha\beta}$ is the surface tension of the corresponding
interface. It was assumed that complete wetting by the $\alpha$
phase occurs at the W-$\beta$ interface. For a detailed discussion of 
the Kelvin equation in the context of liquid crystal phase transitions 
see Ref. \cite{Sluckin}.  
Applying this equation using $H/\sigma=28.88$ (the location of the 
minimum in the $\mu-H$ phase diagram of Fig. \ref{fig12}), we obtain 
an undersaturation $\beta\Delta\mu=-0.0429$, while its real value 
is $\beta\Delta\mu=-0.1255$. In the derivation of the Kelvin equation, 
deviations from the bulk structure of the density profile arising
from the confinement by the external potential are neglected. Also, the elastic 
energy resulting from the compression or expansion of a layered phase 
confined between two walls is not taken into account. These effects might be 
responsible for the differences found between our calculations and     
the estimation based on the Kelvin equation.
We have checked that the sequence of minima in the $\mu-H$ 
phase diagram tends to $\mu_0$ as $H\to\infty$, a 
result predicted by Eq. (\ref{Kelvin}).  

\begin{figure}[h]
\epsfig{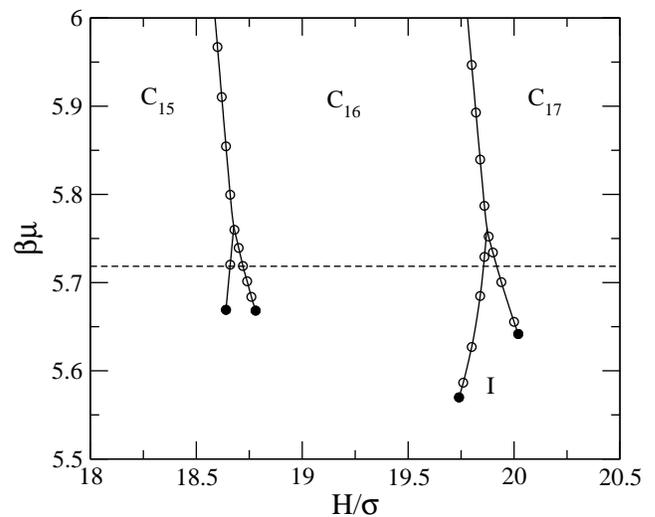}
\caption{$\mu-H$ surface phase diagram showing critical points 
(filled circles). Number of columnar layers are indicated as subscripts.}
\label{fig13}
\end{figure}

Refs. \cite{Roij} and \cite{Dani} showed that the capillary nematization 
line of the confined liquid-crystal fluid ends in a critical point 
for small values of the pore width. In order to
study how the topology of the surface phase diagram changes 
in the regime of small pore widths, we have 
carried out the corresponding calculations of interfacial structure.
We have found that the I-C capillary ordering transition changes at some 
particular value of $H$ (near its maximum undersaturation represented
by the minimum in the I-C interface coexisting curve) 
from first to second order. For lower values of $H$ two critical 
points emerge from this single point, the distance between them
increasing. In Fig. \ref{fig13} one of these scenarios is shown. As we 
can see, there is a range of values of $H$ (near the triple points) 
where the first-order capillary ordering transitions are still present
but, between the critical points belonging to different layering branches, 
columnar ordering grows continously from the 
isotropic (damped columnar interfacial phase) 
to a highly-ordered columnar phase. Layering transitions 
are always present even for very small $H$, as will be shown below.  
An interesting feature of this phase diagram is that the location
of the triple points moves above the bulk coexistence value $\mu_0$.
This indicates that the interfacial columnar phase just below the
triple points can be unstable for values
of chemical potentials corresponding to those of
columnar-phase stability at bulk
(similar to the capillary evaporation of the confined fluid).
For wide enough slits (those for which the parabolas are connected)
the triple points are practically located at $\mu_0$, as can be
observed from Fig. \ref{fig12}. 

\begin{figure}[h]
\epsfig{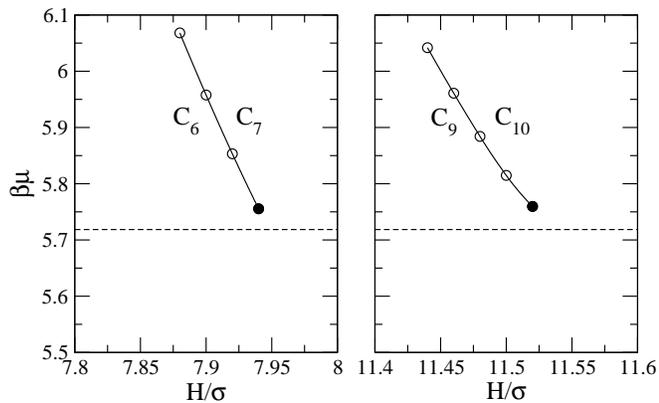}
\caption{$\mu-H$ surface phase diagrams for small values of $H$.} 
\label{fig14}
\end{figure}

For even smaller values of $H$, only layering transitions 
remain; these end in critical points located above $\mu_0$, as 
Fig. \ref{fig14} shows. 

\begin{figure}[h]
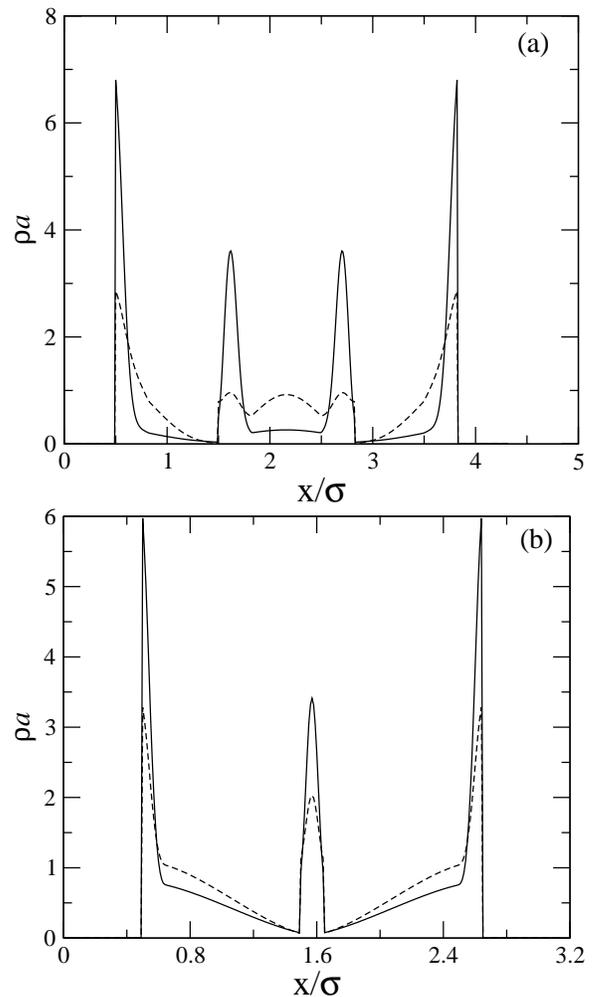

\epsfig{file=fig15a.eps,width=3.0in}
\epsfig{file=fig15b.eps,width=3.0in}
\caption{(a): density profiles of two coexisting phases (shown with 
solid and dashed lines)
at $\beta\Delta\mu=0.5176$. The pore width is 
$H/\sigma=4.32$.  
(b): same as in (a) but for a pore with $H/\sigma=3.14$ and 
for $\beta\Delta\mu=0.72836$.}   
\label{fig15}
\end{figure}

When the width $H$ is such that the pore can only accommodate one 
particle with its long axis perpendicular to the wall (or not more 
than four or three particles aligned parallel to the wall) the system is near 
the one-dimensional limit. It is known that hard-core systems in this 
limit do not exhibit first-order phase transitions, but even for 
very narrow slits we can still find first-order transitions at which
the density profile experiences an abrupt change inside the pore. 
In Fig. \ref{fig15} (a) and (b)
we show two coexisting density profiles corresponding to  
oversaturations, $\beta\Delta\mu=0.51760$ and 
$\beta\Delta\mu=0.72836$, and pore widths $H/\sigma=4.32$ and 
$H/\sigma=3.14$ 
in (a) and (b), respectively. The fluid inside the pore undergoes
a phase transition, which dramatically changes the structure of the interfacial 
density profiles by increasing the heigh of four [Fig. \ref{fig15}(a)] 
or three [(b)] density peaks inside the pore.

\section{Conclusions}
In this article we have shown that 2D fluids composed 
of anisotropic particles 
interacting via hard-core repulsion and confined in a slit geometry
exhibit a complex and rich interfacial phase behavior. Apart from 
first-order capillary columnar ordering, we have also found layering 
transitions in this system. These results are similar to those found 
in 3D liquid-crystal fluids confined in a pore, where capillary 
smectization and layering phenomena were also found \cite{Dani}. 
In view of these 
similarities, we can extract the conclusion that, independent of
the system dimensionality and the peculiarities of the layered phases, 
either smectic or columnar, if the fluid-wall interaction enhances 
layered interface ordering (homeotropic in case of smectic phases, 
and the entropically favored parallel alignment for the columnar phase), 
compatible with the equilibrium bulk phase, then the confined 
fluid exhibits the interfacial phase transitions described above.

In this study we have used as a model a hard-rectangle fluid, and 
the density and the order-parameter profiles were calculated by minimizing 
the excess surface free-energy functional resulting from the 
Fundamental-Measure Theory applied to the two-dimensional Zwanzig model.  
The orientational degrees of freedom were discretized, in order to
take advantage of having a free-energy functional which reduces to the 
exact one-dimensional functional when the density profile is constrained 
to lie along a line. This property is crucial to study strongly
confined fluids (as is the case in this study), in particular when the 
pore width has only a few particle diameters in width.

As already pointed out in Sec. \ref{intro},
some experiments had shown
profound similarities between particle configurations
obtained as stationary states of systems of anisotropic grains
and those corresponding to the equilibrium states obtained by density
functional minimization \cite{Narayan}. These similarities can be explained by
applying a maximum-entropy principle on granular collections of particles, i.e.
for a fixed packing fraction, externally-induced vibrational
motion forces the system to explore those stationary states which maximize
the configurational entropy (since the grains cannot overlap).
Of course, equilibrium statistical mechanics is unable to propose an equation
of state for granular matter, but it could be possible to predict
that granular matter composed of anisotropic particles and confined between
parallel walls may support a stationary texture consisting of layers
of particles oriented parallel to the wall. The manner in which
the grain orientations propagate into the container would depend on the
average packing fraction and on the frequency of the external force.
Only at this qualitative level can we give some insight into possible
complete wetting phenomena and capillary ordering in
granular rod fluids confined between two horizontal plates at a distance
slightly larger than the particle dimensions in the vertical direction
(thus simulating a two-dimensional system), and also confined by one or two
vertical planes (these playing the role of hard walls).

Some calculations (not shown here) on the 2D HR fluid show that, for different
aspect ratios, 2D smectic and crystal phases can be stable over some
range of packing fractions. It would be interesting to explore whether 
confinement suppresses or enhances bulk ordering, and to study the 
changes in the surface phase diagram when phases of different symmetries
are included. Work along this direction is currently in progress.

\section*{Acknowledgments}
I thank D. de las Heras, E. Velasco, and L. Mederos for
useful discussions, and E. Velasco for a critical reading of the 
manuscript.
The author gratefully acknowledges financial support from Ministerio de
Educaci\'{o}n y Ciencia
under grants No. 
BFM2003-0180 and from Comunidad Aut\'onoma de Madrid
(S-0505/ESP-0299) and (UC3M-FI-05-007). The author was supported by a
Ram\'on y Cajal research contract from the Ministerio de Educaci\'on
y Ciencia.


\begin{thebibliography}{13}

\bibitem{Schmidt} M. Schmidt, and H. L\"owen, Phys. Rev. Lett. {\bf 76}, 
4552 (1996); Phys. Rev. E {\bf 55}, 7228 (1997). 
\bibitem{Dijkstra} M. Dijkstra, Phys. Rev. Lett. {\bf 93}, 108303 
(2004); A. Fortini, and M. Dijkstra, J. Phys. Condes Matter {\bf 18}, 
L371 (2006).
\bibitem{Binder} L. Salamacha, A. Patrykiejew, S. Sokolowski, and K. 
Binder, Eur. Phys. J. E {\bf 13}, 261 (2004); 
L. Salamacha, A. Patrykiejew, S. Sokolowski, Eur. Phys. J. E {\bf 18}, 
425 (2005).
\bibitem{Roij} R. van Roij, M. Dijkstra, and R. Evans, 
Europhys. Lett. {\bf 49}, 350 (2000); M. Dijkstra, R. van Roij, and 
R. Evans, Phys.Rev. E {\bf 63}, 051703 (2001)
\bibitem{Harnau} L. Harnau, and S. Dietrich, Phys. Rev. E {\bf 66}, 
051702 (2002).
\bibitem{Kike} I. Rodriguez-Ponce, J. M. Romero-Enrique, E. Velasco, 
L. Mederos, and L. F. Rull, J. Phys.: Condens. Matter {\bf 12}, 
A363 (2000); I. Rodriguez-Ponce, J. M. Romero-Enrique, and L. F. 
Rull, Phys. Rev. E {\bf 64}, 051704 (2001)
\bibitem{Samo} R. E. Webster, N. J. Mottram, and 
D. J. Cleaver, Phys. Rev. E. {\bf 68}, 021706 (2003); 
Z. Kutnjak, S. Kralj, G. Lahajnar, and S. Zumer, Phys. Rev. E 
{\bf 70}, 051703 (2004).
\bibitem{Dani} D. de las Heras, E. Velasco, and L. Mederos, 
Phys. Rev. Lett. {\bf 94}, 017801 (2005); Phys. Rev. E {\bf 74}, 011709 
(2006).
\bibitem{Kaganer} V. M. Kaganer, H. Mohwald, and P. Dutta, 
Rev. Mod. Phys. {\bf 71}, 779 (1999).
\bibitem{Aranson}I. S. Aranson, and L. S. Tsimring, Rev. Mod. Phys. 
{\bf 78}, 641 (2006).
\bibitem{Reis} P. M. Reis, R. A. Ingale, and M. D. Shattuck, 
Phys. Rev. Lett. {\bf 96}, 258001 (2006).
\bibitem{Narayan} V. Narayan, N. Menon, and S. Ramaswamy, 
J. Stat. Mech. P01005 (2006)
\bibitem{Yuri1} Y. Mart\'{\i}nez-Rat\'on, E. Velasco, and L. Mederos, 
J. Chem. Phys. {\bf 122}, 064903 (2005); J. Chem. Phys. {\bf 125}, 
014501 (2006).
\bibitem{Donev} A. Donev, J. Burton, F. H. Stillinger, and 
S. Torquato, Phys. Rev. B {\bf 73}, 054109 (2006).
\bibitem{Coniglio} A. Fierro, M. Nicodemi, and A. Coniglio, 
Eurphys. Lett. {\bf 59}, 642 (2002); Phys. Rev. E {\bf 66}, 
061301 (2002); A. Coniglio, A. Fierro, and N. Nicodemi, 
Eur. Phys. J. E {\bf 9}, 219 (2002).
\bibitem{Dean} D. S. Dean, and A. Lefevre, Phys. Rev. Lett {\bf 90}, 
198301 (2003).
\bibitem{Galanis} J. Galanis, D. Harries, and D. L. Sackket, 
Phys. Rev. Lett. {\bf 96}, 028002 (2006). 
\bibitem{Chaudhuri} D. Chaudhuri, and S. Sengupta, Phys. Rev. Lett. 
{\bf 93}, 115702 (2004).
\bibitem{Yasha}Y. Rosenfeld, M. Schmidt, H. L\"owen, and P. Tarazona, 
J. Phys.: Condens. Matt. {\bf 8}, L577 (1996); Phys. Rev. E {\bf 55}, 
4245 (1997); P. Tarazona, and Y. Rosenfeld, Phys. Rev. E {\bf 55}, 
R4873 (1997).
\bibitem{Yuri} J. A. Cuesta and Y. Mart\'{\i}nez--Rat\'on, Phys. Rev. Lett.
{\bf 78}, 3681 (1997); J. Chem. Phys. {\bf 107}, 6379 (1997).
\bibitem{Roij2} R. van Roij, P. Bolhuis, B. Mulder, and 
D. Frenkel, Phys. Rev. E {\bf 52}, R1277 (1995).
\bibitem{Allen} J. S. van Duijneveldt, and M. P. Allen, Molec. Phys. {\bf 90}, 
243 (1997).
\bibitem{PRE_Yuri} Y. Mart\'{\i}nez-Rat\'on, Phys. Rev. E 
{\bf 69}, 061712 (2004).
\bibitem{Lipowsky} R. Lipowsky, J. Appl. Phys. {\bf 55}, 2485 (1984).
\bibitem{Holyst} A. Poniewierski, and R. Holyst, 
Phys. Rev. A {\bf 38}, 3721 (1988).
\bibitem{Chrzanowska} A. Chrzanowska, P. I. C. Teixeira, H. Ehrentraut, 
and D. J. Cleaver, J. Phys.: Condens. Matter 
{\bf 13}, 4715 (2001).
\bibitem{Yuri2}Y. Mart\'{\i}nez-Rat\'on, A. M. Somoza, L. Mederos, and 
D. E. Sullivan, Faraday Discuss. {\bf 104}, 111 (1996); Y. Martinez, 
A. M. Somoza, L. Mederos, and D. E. Sullivan, Phys. Rev. E
{\bf 53} 2466 (1996).
\bibitem{Sluckin} T. J. Sluckin and A. Poniewierski, Molec. Cryst. Liq. Cryst. 
{\bf 179}, 349 (1990).
\end{thebibliography}
\end{document}